\documentclass[aps,amsmath,amssymb,prl,twocolumn,superscriptaddress]{revtex4}
\usepackage{graphicx}
\usepackage{textcomp}
\usepackage{gensymb}
\usepackage{xcolor}
\usepackage[normalem]{ulem}

\newcommand{\comment}[1]{}

\newcommand{\Gedge}{G_\mathrm{edge}}

\newcommand{\sigb}{\sigma_\mathrm{bulk}}
\newcommand{\sige}{\sigma_\mathrm{edge}}
\newcommand{\Vmax}{V_g^\mathrm{max}}
\newcommand{\Rmax}{\rho_{xx}^\mathrm{max}}
\comment{
2010: Knez et al observe a finite bulk conductivity in the inverted regime which they attribute to the theory of Laikhman.

2011: knez et al claim they observe edge states. This is basically an extrapolation of the conductance. The bulk conductivity is high.

2013: Suzuki et al observe edge states in the opoological phases (RNL) with a low residual bulk G due to Be doping

2014: knez et al. claim they observe edge states in macroscopic devices. Si-doped DQW.

2015: Du et al (houston) introduce  Si doping. They observe thermal activation and localization in Rxx, and some quantized plateaus.

2016: Michele observe trivial edge states.

2016: Karalic see beating in the LL in the VB, attributed to strong spin-orbit.

2016: Nguien et al.

2019: using Si doped DQW Sazgari demonstrate trivial edge state conductance with localization due to disorder.

2020: Irie et al obtain large gap and large resistivity using strain only. Largest gap: 35 meV mechainsm for edge conduction : fermi level pinning (Michele2016)

note (nhye,: lowering mobility by intentionally introducing disorder, is accepted as a mean to suppress bulk conductivity)
}

\begin{document}


\title{Large inverted band-gap in strained three-layer InAs/GaInSb quantum wells}

\author{C.~Avogadri}
\affiliation{Laboratoire Charles Coulomb (L2C), UMR 5221 CNRS-Universit\'e de Montpellier, Montpellier, France.}

\author{S.~Gebert}
\affiliation{Laboratoire Charles Coulomb (L2C), UMR 5221 CNRS-Universit\'e de Montpellier, Montpellier, France.}
\affiliation{Institut d'Electronique et des Syst\`emes (IES), UMR 5214 CNRS-Universit\'e de Montpellier, Montpellier, France.}

\author{S.~S.~Krishtopenko}
\affiliation{Laboratoire Charles Coulomb (L2C), UMR 5221 CNRS-Universit\'e de Montpellier, Montpellier, France.}

\author{I.~Castillo}
\affiliation{Laboratoire Charles Coulomb (L2C), UMR 5221 CNRS-Universit\'e de Montpellier, Montpellier, France.}

\author{C.~Consejo}
\affiliation{Laboratoire Charles Coulomb (L2C), UMR 5221 CNRS-Universit\'e de Montpellier, Montpellier, France.}

\author{S.~Ruffenach}
\affiliation{Laboratoire Charles Coulomb (L2C), UMR 5221 CNRS-Universit\'e de Montpellier, Montpellier, France.}

\author{C.~Roblin}
\affiliation{Laboratoire Charles Coulomb (L2C), UMR 5221 CNRS-Universit\'e de Montpellier, Montpellier, France.}

\author{C.~Bray}
\affiliation{Laboratoire Charles Coulomb (L2C), UMR 5221 CNRS-Universit\'e de Montpellier, Montpellier, France.}

\author{Y.~Krupko}
\affiliation{Laboratoire Charles Coulomb (L2C), UMR 5221 CNRS-Universit\'e de Montpellier, Montpellier, France.}

\author{S.~Juillaguet}
\affiliation{Laboratoire Charles Coulomb (L2C), UMR 5221 CNRS-Universit\'e de Montpellier, Montpellier, France.}

\author{S.~Contreras}
\affiliation{Laboratoire Charles Coulomb (L2C), UMR 5221 CNRS-Universit\'e de Montpellier, Montpellier, France.}

\author{A.~Wolf}
\affiliation{Technische Physik, Physikalisches Institut and W\"urzburg‐Dresden Cluster of Excellence ct.qmat, Am Hubland, D-97074 W\"urzburg, Germany}

\author{F.~Hartmann}
\affiliation{Technische Physik, Physikalisches Institut and W\"urzburg‐Dresden Cluster of Excellence ct.qmat, Am Hubland, D-97074 W\"urzburg, Germany}

\author{S.~H\"ofling}
\affiliation{Technische Physik, Physikalisches Institut and W\"urzburg‐Dresden Cluster of Excellence ct.qmat, Am Hubland, D-97074 W\"urzburg, Germany}

\author{G. Boissier}
\affiliation{Institut d'Electronique et des Syst\`emes (IES), UMR 5214 CNRS-Universit\'e de Montpellier, Montpellier, France.}

\author{J.-B. Rodriguez}
\affiliation{Institut d'Electronique et des Syst\`emes (IES), UMR 5214 CNRS-Universit\'e de Montpellier, Montpellier, France.}

\author{S.~Nanot}
\affiliation{Laboratoire Charles Coulomb (L2C), UMR 5221 CNRS-Universit\'e de Montpellier, Montpellier, France.}

\author{E.~Tourni\'e}
\affiliation{Institut d'Electronique et des Syst\`emes (IES), UMR 5214 CNRS-Universit\'e de Montpellier, Montpellier, France.}

\author{F.~Teppe}
\email[]{frederic.teppe@umontpellier.fr}
\affiliation{Laboratoire Charles Coulomb (L2C), UMR 5221 CNRS-Universit\'e de Montpellier, Montpellier, France.}

\author{B.~Jouault}
\email[]{benoit.jouault@umontpellier.fr}
\affiliation{Laboratoire Charles Coulomb (L2C), UMR 5221 CNRS-Universit\'e de Montpellier, Montpellier, France.}

\begin{abstract}
Quantum spin Hall insulators (QSHIs) based on HgTe and three-layer InAs/GaSb quantum wells (QWs) have comparable bulk band-gaps of about 10--18~meV. The former however features a band-gap vanishing with temperature, while the gap in InAs/GaSb QSHIs is rather temperature-independent.
Here, we report on the realization of large inverted band-gap in strained three-layer InAs/GaInSb QWs. By temperature-dependent magnetotransport measurements of gated Hall bar devices, we extract
a gap as high as 45 meV. Combining local and non-local measurements, we attribute the edge conductivity observed at temperatures up to 40 K to the topological edge channels with equilibration lengths of a few micrometers. Our findings pave the way toward manipulating edge transport at high temperatures in QW heterostructures.
\end{abstract}

\maketitle


\emph{Introduction.}-- Time-reversal invariant two dimensional (2D) topological insulators, also known as quantum spin Hall insulators (QSHIs), are characterized by insulating bulk
and spin-polarized topologically protected states at the sample edges~\cite{2005Kane,2006Bernevig}. The presence of these edge states is of great interest for potential applications in spintronics, metrology~\cite{2019Yahniuk} and quantum information~\cite{2010Hasan,2011Qi}. So far, the QSHI state was experimentally established in HgTe quantum wells (QWs)~\cite{2007Konig}, InAs/GaSb QW bilayers~\cite{2008Liu,2011Knez} and 1$T'$-WTe$_2$ monolayers~\cite{q1}. The latter with its 45~meV inverted band-gap~\cite{q2,q3} demonstrated a stable QSHI state up to 100~K. This motivates the search of other 2D systems with even wider inverted band-gaps, but the observation of QSHI in monolayer systems is experimentally challenging because of structural or chemical instabilities~\cite{q4,q5,q6} and non-mature technological processing. This stimulates the search of alternative candidates for high-temperature QSHI among QW heterostructures.

The first measurements of the quantized edge conductance (main characteristic of QSHI) in HgTe QWs were performed at temperatures in the range of a few tens of mK~\cite{2007Konig}. Indeed, a relatively small inverted band-gap (typically lower than 15~meV) in HgTe QWs grown on CdTe buffer makes it difficult to observe the quantized edge conductance at elevated temperatures. Note that strain engineering using virtual substrate increases the band-gap up to 55 meV in compressively strained QWs~\cite{q7}. Such high values, however, occur at low temperatures only, whereas increasing temperature yields the band-gap vanishing and topological phase transition into trivial state~\cite{q10,2016Krishtopenko,q10b,q8,q9}. Hence, the observation of the QSHI state in HgTe Qws is so far limited to 15~K~\cite{q11}.

QSHIs based on InAs/GaSb QW bilayers raise a considerable interest over HgTe QWs due to their ease of fabrication. However, their small inverted band-gap of about 3--4~meV induces a large residual bulk conductance~\cite{2001Naveh}. This limits the observation of quantized edge conductance values to the millikelvin temperature range~\cite{2011Knez,2012Knez,2013Du,2015Du}. Although the residual bulk conductance can be indeed reduced by means of various techniques (implantation of Si impurities at the InAs/GaSb interface~\cite{2014Knez}, Be doping~\cite{2015Suzuki}, or the use of low-purity Ga source for MBE growth~\cite{2013Charpentier}), the quantized values of edge conductance out of the millikelvin range have not yet been  observed even in strained InAs/GaInSb QW bilayers with a higher band-gap~\cite{2016Akiho,2017Dua,2020Irie}.

Removing the structure inversion asymmetry inherent to InAs/GaSb QW bilayers by adding a second InAs layer significantly enhances the inverted band-gap energy~\cite{2018KrishtopenkoScience}, resulting in QSHI with the bulk gap comparable with the one of inverted HgTe QWs. Despite the general similarities and characteristics of topological states in HgTe~QWs~\cite{2016Krishtopenko} and three-layer (3L) InAs/GaSb~(QWs)~\cite{2018KrishtopenkoScience,q12,2019Krishtopenko,2019Krishtopenkoa}, the inverted band-gap of the latter is rather temperature-independent~\cite{2018Krishtopenko,2021Meyer}.
This fact, as well as the theoretically predicted inverted band-gap in strained 3L InAs/GaInSb~QWs above 60~meV~\cite{2018KrishtopenkoScience}, make these QWs extremely attractive for observing quantized edge conductance at high temperatures.

This work reports on the experimental realization of strained 3L InAs/GaInSb~QWs with large inverted band-gap. Temperature-dependent magnetotransport measurements of Hall bar devices made from 3L InAs/Ga$_{0.65}$In$_{0.35}$Sb QWs in local and non-local geometries reveal energy gaps as high as 45 meV, associated with edge conductance attributed to topological states. Note that the experimental gap values can be further enhanced by growing 3L InAs/Ga$_{1-x}$In$_{x}$Sb QWs with higher values of $x$ (cf.~Ref.~\onlinecite{2020Irie}).

\begin{table*}
\caption{Main sample parameters.
For all fabricated Hall bars, $W$ is the width, $\ell_p$ is the distance between the lateral probes and $l_1$ is the distance between the source (or drain) contact and the closest lateral probe. }
\label{table:samples}
\begin{tabular}{@{}c|c|c|c|c|c|c|c@{}}
\hline
\begin{tabular}[c]{@{}l@{}}Sample\end{tabular} & \begin{tabular}[c]{@{}l@{}}InAs \\ thickness~(nm)\end{tabular} & \begin{tabular}[c]{@{}l@{}}Ga$_{1-x}$In$_{x}$Sb \\ thickness~(nm)\end{tabular} & \begin{tabular}[c]{@{}l@{}}Outer barriers\end{tabular} & \begin{tabular}[c]{@{}l@{}}Metamorphic \\ buffer \end{tabular}  &  \begin{tabular}[c]{@{}l@{}}Device index ($W$, $\ell_p$, $l_1$ in $\mu$m)\end{tabular} & \begin{tabular}[c]{@{}l@{}}$\Delta$ theory \\ (meV)\end{tabular} & \begin{tabular}[c]{@{}l@{}}$\Delta$ exp.\\  (meV)\end{tabular} \\ \hline
S3054    & 10.3       & 4.3 ($x=0.00$) & Al$_{0.9}$Ga$_{0.1}$As$_{0.07}$Sb$_{0.93}$  & GaSb      &  HB0$(10,22,17)$ & 14   & N/A      \\
S3052    & 7.5        & 3.1 ($x=0.35$) & Al$_{0.9}$Ga$_{0.1}$As$_{0.07}$Sb$_{0.93}$ & AlSb         &  HB1$(100,220,170)$ & 30   & $30\pm2$     \\
S3198    & 7.5        & 3.1 ($x=0.35$) & AlSb   & AlSb  &  HB4$(20,10,40)$, HB6$(20,30,30)$& 45   & $45\pm8$ \\ \hline
\end{tabular}
\end{table*}

\begin{figure}
	\includegraphics[width=0.95 \linewidth]{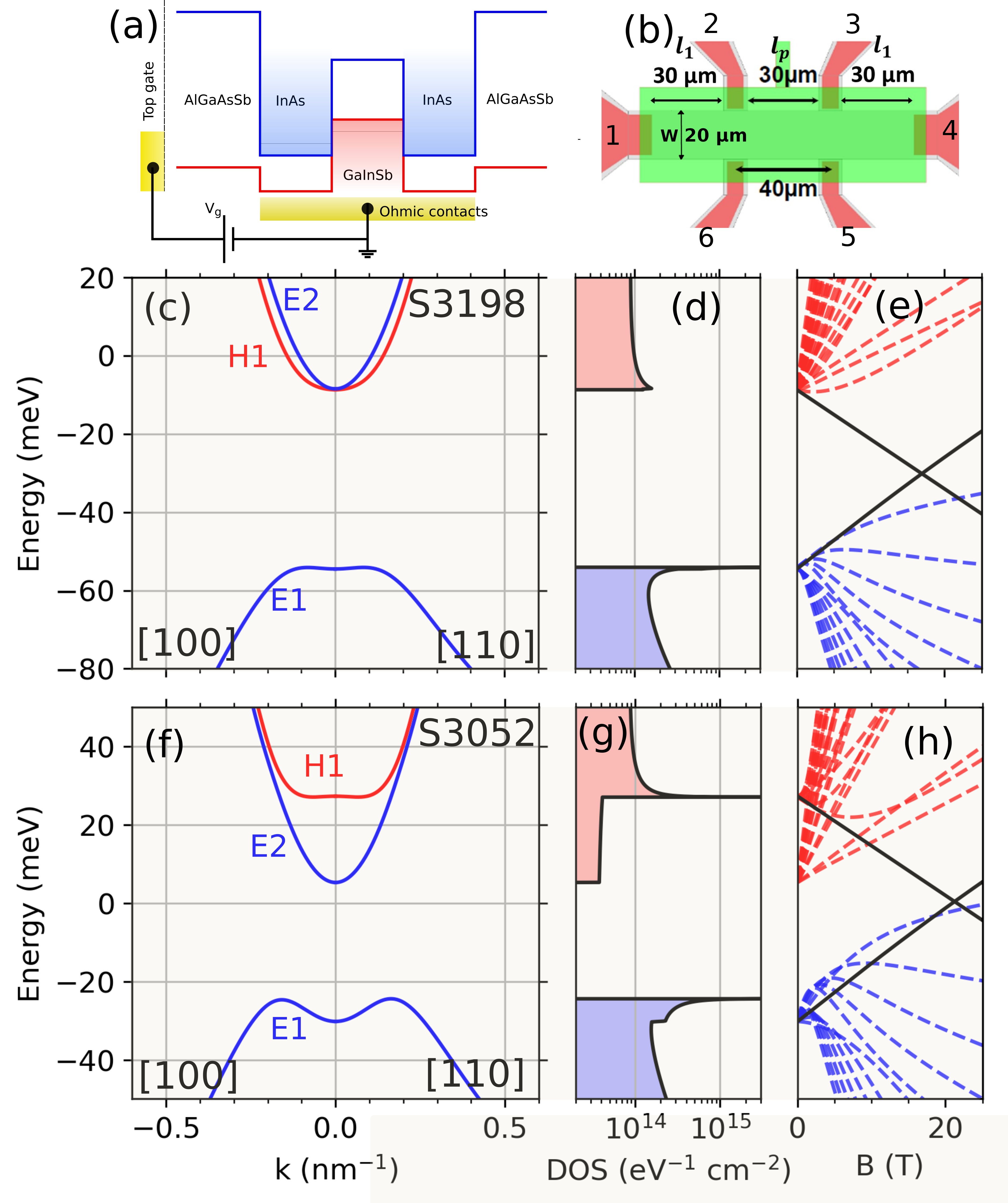}
	\caption{
		(a) Qualitative scheme of 3L QWs.
		(b) Sketch of the Hall bar HB6. (c--g) Band structure, density of states and Landau levels for samples S3198 (c,d,e) and S3052 (e,f,g). The zero-mode Landau levels~\cite{2007Konig,2018KrishtopenkoScience} in panels (e,h) are indicated by black solid lines.}
	\label{fig:intro}
\end{figure}

\emph{Materials and methods.}--We have fabricated a set of three QWs: S3054, S3052 and S3198, with distinct strain and thickness parameters (see Table~\ref{table:samples}). The samples were grown by molecular beam epitaxy. The active part of the samples sketched in Fig.~\ref{fig:intro}(a) contains a symmetric 3L InAs/Ga$_{1-x}$In$_x$Sb QW embedded between AlGaAsSb barriers. For the samples S3052 and S3198, the widths of the InAs and Ga$_{0.65}$In$_{0.35}$Sb layers were chosen to be 25 and 10 monolayers, respectively. For the sample S3054, the width of the InAs and GaSb layers were increased up to 34 and 14 monolayers. Both samples S3054 and S3052 were grown on semi-insulating GaAs~(001) substrates, whereas sample S3198 was grown on a GaSb~(001) substrate.
The samples have also different metamorphic buffer layers, and different strain states.

All the samples were processed by optical lithography into micro-sized Hall bar devices with a metallic front gate, on a plasma-enhanced chemical vapor deposited (PECVD) 300~nm-thick SiO$_2$ for samples S3052 and on a 110-nm-thick stacking of SiO$_2$/Si$_3$N$_4$ dielectric insulators for S3054 and S3198. Transport measurements of various gated Hall bars (see Tab.~\ref{table:samples}) were performed in a cryostat equipped either with a variable temperature insert or with a Helium-3 insert for the temperatures below 1.7~K. We used standard lock-in measurements with 10~nA currents at 11 Hz and high-impedance 1~T$\Omega$  preamplifiers. The quantity $R_{ij,kl}$ corresponds to the voltage between the probes $k$ and $l$ divided by the current flowing between contacts $i$ and $j$.

\emph{Bulk band-gap.}-- Figures~\ref{fig:intro}(c,f) represent realistic band structure calculations~\cite{2018KrishtopenkoScience}. 
All the samples have an inverted band structure with the hole-like $H1$ band lying above the electron-like $E1$ band~\cite{SM}.
The calculated band-gap for the samples S3052 and S3198 is $\Delta \simeq 30$~meV and $45$~meV, respectively. Here, we note an influence of the outer barriers on the band structure of 3L InAs/GaInSb QWs at given layer thicknesses. The sample S3054 with the smaller gap ($\Delta \simeq 14$~meV) is similar to the one studied in Ref.~\onlinecite{2018Krishtopenko}, in which the inverted band structure was evidenced.

\begin{figure}
	\includegraphics[width=0.95 \linewidth]{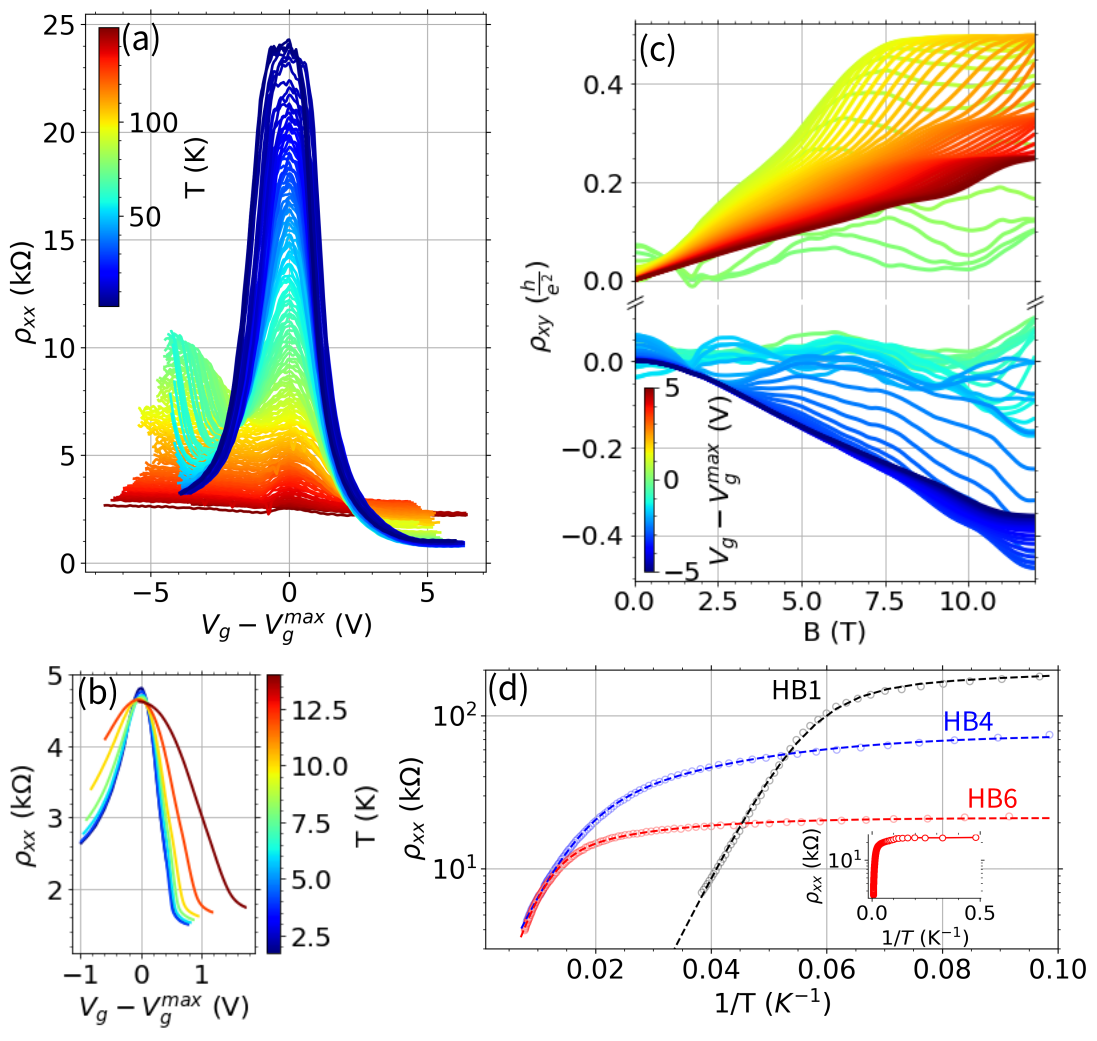}
	\caption{(a,b) Longitudinal resistivity $\rho_{xx}(V_g)$ at different temperatures for HB6~(a) and HB0 device~(b). (c)~Transverse magnetoresistance $\rho_{xy}(B)$ as a function of $V_g$ at $T=300$~mK for HB6 device. For clarity, a vertical voltage offset has been introduced between $V> \Vmax$ and $V\leq \Vmax$. (d) Temperature dependence of the peak resistivity $\rho_{xx}^\mathrm{max}$ for the three devices: HB1 (S3052), HB4 (S3198) and HB6 (S3198).
Open symbols corresponds to the experimental values, while the dashed curves represent the fits as indicated in the text. The inset shows the saturation of $\rho_{xx}^\mathrm{max}$ for HB6 device at lower temperatures down to 2~K.}
	\label{fig:gap}
\end{figure}

We first focus on measuring the longitudinal resistivity 
$\rho_{xx} \simeq R_{14,23}W/\ell_p$ 
as a function of the gate voltage $V_g$. In HB0 device, $\rho_{xx}(V_g)$ displays a peak at $T=2$~K,
indicating a gap opening. However, this peak culminates at $\rho_{xx} \simeq $ 4.8~k$\Omega$ only with a weak insulating behavior as shown in Fig.~\ref{fig:gap}(b). Thus, the band-gap in HB0 device is shunted either by edge states or a parasitic conductivity channel in one of the cap or buffer layers~\cite{SM}. Further, we focus mainly on HB6 device (S3198 sample).
Figure~\ref{fig:gap}(a) shows $\rho_{xx}(V_g)$ for HB6 device, which evidences a much higher peak of around 25~k$\Omega$ at low temperature. For clarity, each curve has been horizontally offset by the position of the peak maximum $V_g^\mathrm{max}$. The main peak is flanked on its left side by a dip around $V_g-V_g^\mathrm{max}= -2.5$~V. Such local $\rho_{xx}$ minimum can be attributed to the van Hove singularity (VHS) at the top of the valence band as seen in Fig.~\ref{fig:intro}(d). Similar dips were also observed in InAs/GaSb QW bilayers~\cite{1997Yang,2012Brihuega,2016Karalic}, and recently in 3L InAs/GaSb QWs~\cite{2021Meyer}.

Figure~\ref{fig:gap}(c) shows the transverse magnetoresistance $\rho_{xy}$ for HB6 device at $T=300$~mK. An ambipolar behavior centered at $V_g=\Vmax$ is evident. In the  conduction band (CB), $\rho_{xy}$ is linear at low $B$ with a pronounced quantum Hall effect at high magnetic field. In the valence band (VB), $\rho_{xx}$ is bended below $B=2$~T even at the lowest available gate voltage, $V_g-\Vmax=-5$~V. At this voltage, both longitudinal and transverse magnetoresistances are satisfactorily fitted by taking into account two types of carriers: hole-like carriers of density $n_h=7.0\times 10^{11}$ cm$^{-2}$ and mobility $\mu_h \simeq 1,000$~cm$^2$/V$\cdot$s, and electron-like carriers of density $n_e=0.4\times 10^{11}$~cm$^{-2}$ and mobility $\mu_e \simeq 10,000$ cm$^2$/V$\cdot$s. This agrees well with the band structure near the top of the VB, where the Fermi surface has two distinct contours: an inner contour representing electron-like particles and an outer contour corresponding to hole-like particles (see Fig.~\ref{fig:intro}(c)). In accordance with the band structure calculations shown in  Fig.~\ref{fig:intro}(e), the occupied low-indices Landau levels at $B>2$~T are formed by the outer contour states, while the states of the inner contour become depopulated. Experimentally, the measurements of Shubnikov-de Haas (SdH) oscillations at $V_g-\Vmax=-5$~V reveal a single frequency above $B=5$~T corresponding to the carrier concentration $n_{SdH}=8.3\times 10^{11}$~cm$^{-2}$. This value corresponds roughly to $n_h+n_e\simeq 7.4\times 10^{11}$~cm$^{-2}$, as obtained from the low-field analysis.

Figure~\ref{fig:gap}(d) summarizes the temperature dependance of the peak value of the resistivity,  $\rho_{xx}^{\mathrm{max}}$, 
for three Hall bars devices: HB1 (S3052), HB4 and HB6 (S3198). At high temperatures (above 25~K for S3052 and 150~K for S3198), the samples demonstrate an additional planar conduction, therefore the corresponding data were discarded. At lower temperatures, a strong increase of $\rho_{xx}^\mathrm{max}$ was observed, followed  by a weaker temperature dependance at even lower $T$. The latter is typically attributed to disorder-induced localization gap or edge states, while the strong temperature dependence is associated with thermal activation through the band-gap. Note that a similar behavior of $\rho_{xx}^{\mathrm{max}}(T)$ has been commonly observed in inverted InAs/Ga(In)Sb QW bilayers~\cite{2010Knez,2013Suzuki,2015Du,2020Irie}.

As seen from Fig.~\ref{fig:gap}(d), $\rho_{xx}^{\mathrm{max}}$ as a function of $T$ is well fitted by the sum of two activation processes with an additional constant term~\cite{2020Irie}:
$\left(\rho_{xx}^\mathrm{max}\right)^{-1}=\sigma_a\exp{\left(-{\Delta}/{2k_BT}\right)}
+\sigma_\mathrm{loc}\exp{\left(-{\Delta_\mathrm{loc}}/{k_B T}\right)+\sigma_0}$,
where $k_B$ is the Boltzmann constant and $\sigma_0$, $\sigma_a$, $\sigma_\mathrm{loc}$
($\Delta$ and $\Delta_\mathrm{loc}$) have the dimensions of conductivity (energy).
As we impose $\Delta_\mathrm{loc}<\Delta$, the term $\sigma_a\exp{\left(-{\Delta}/{2k_BT}\right)}$
represents the strong $T$-dependence, while the two other terms describe weak temperature-dependence in the saturation regime. The fits give the energies $\Delta=45\pm 8$ meV for S3198 (HB4 and HB6), $\Delta=30\pm 2$ meV for S3052 (HB1), and $\Delta_\mathrm{loc}$ lying in the range 0.7--1.2~meV for all devices (cf. Ref~\onlinecite{2020Irie}). As seen, experimental band-gap energies are in good agreement with their theoretical expectations. Note that the error bar for S3052 is smaller because it does not include the device-to-device variations.

\begin{figure}
	\includegraphics[width=1.0\linewidth]{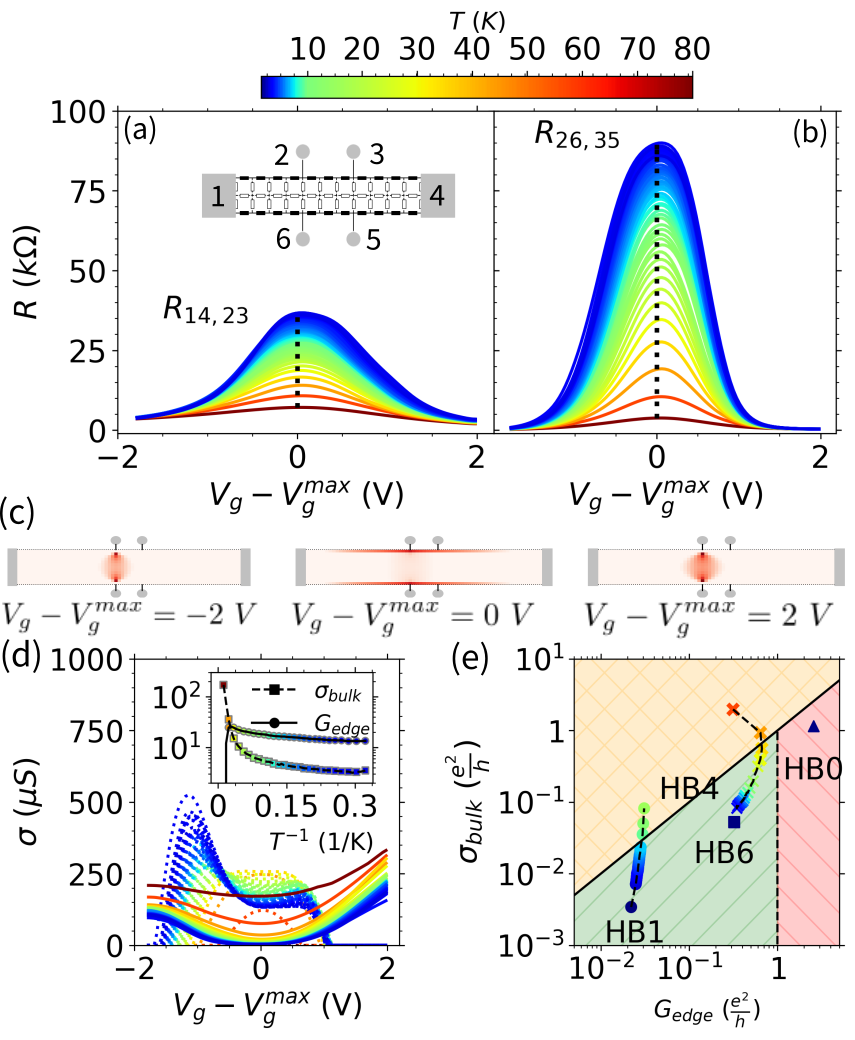}
	\caption{
		(a) Local resistance $R_{14,23}$ and (b) non-local resistance $R_{26,35}$ for HB4 as a function of $V_g-\Vmax$, at different temperatures from $T=80$~K down to $T=3$~K. The inset sketches a minimal  resistor square network. (c) Current dissipation calculated in HB4 at three gate voltages, as given by the fitting of $R_{14,23}$ and $R_{26,35}$ by the network model.
(d) Fitting parameters ($\Gedge \times 10$: dotted lines, $\sigb$: solid lines) extracted from panels (a,b) for HB4 as a function of $V_g-\Vmax$, at different $T$. The inset shows $\Gedge(T)$ and $\sigb(T)$ at $V_g=\Vmax$. (e)  $\Gedge$ and $\sigb$ for HB0 (triangle), HB1 (circle), HB4 (cross), and HB6 (square), at $V_g= \Vmax$ and $T=3$~K. Solid black line: $\sigb= \Gedge$. Vertical dashed line: $\Gedge = e^2/h$ (ballistic edge conduction). Red region: additional parasitic edge conductivity. Green dashed region: diffusive edge conduction. The temperature dependence of ($\Gedge, \sigb$) is indicated for HB1 (up to 16 K) and HB4 (up to 57 K).
}
	\label{fig:maps}
\end{figure}

\emph{Nonlocal resistances.}--
As evidenced from the inset in Fig.~\ref{fig:gap}(d), the resistance peak $\Rmax$ for HB6 device saturates and becomes constant below 10~K, where no activation energy can be found. Further, we demonstrate that this saturation is mainly caused by the conductivity via the edge states.
Figure~\ref{fig:maps}(a) shows the local resistance $R_{14,23}$ as a function of $V_g-\Vmax$ for device HB4 in the temperature range from 3~K up to 80~K. As seen, the peak amplitude is comparable to the one shown in Fig.~\ref{fig:gap}(a), while
the $V_g$ range is reduced in order to focus on the gap region with the insulating behavior.
Figure~\ref{fig:maps}(b) provides the temperature dependence of non-local resistance $R_{26,35}$.
One can see that at $|V_g-\Vmax|\simeq 2$~V corresponding to the edges of CB and VB, the local resistance exceeds the non-local one for all the temperatures. At these $V_g$ values, the non-local resistance finds its origin in the current spreading in the bulk of the Hall bars:
$R_{26,35}\simeq (4\rho/\pi)\exp(-\pi\ell_p/W)$, where $\rho=R_{14,23} W/\ell_p$ is the bulk resistivity.
Experimentally, $R_{26,35}/R_{14,23}\simeq $~0.1--0.3 in the CB and VB~\cite{SM},
which yields $\ell_p \simeq$ 12--17 $\mu$m.
The latter is in qualitative agreement with the geometry of HB4 when the finite width
of the lateral probes (10 $\mu$m) is taken into account.

The situation changes significantly when $V_g$ approaches $\Vmax$ corresponding to the middle of the band-gap. In this case, $R_{26,35}$ increases and becomes more than twice larger than $R_{14,23}$, that cannot be explained within the model above. A similar increase of $R_{26,35}$ was also observed in HB1 and HB6 devices. In HB0 device, even though the non-local resistance increases more moderately, it still becomes ten times higher than the evaluation within the current spreading model.

\emph{Resistive network model.}--
Non-local resistance is often observed in InAs-based QSHI~\cite{2015Du,2015Mueller,2017Mueller,2016Nguyen,2019Sazgari}.
It has been unambiguously related to edge state conduction by combining electric measurements with spatial imaging using SQUID microscopy~\cite{2014Spanton,2016Nichele} and scanning tunneling microscopy~\cite{2019Kaku}. To separate the edge and bulk contributions, each Hall bar device was modeled by a 2D resistor square network, parametrized by the edge and bulk conductivities:
$\sige$ (in $\mu$m/$\ohm$)
and
$\sigb$ (in $\square/\ohm$).
These two parameters are used to fit simultaneously the local and non-local resistances at given gate voltage,
which allows visualizing the resulting current dissipation in the devices, see Fig.~\ref{fig:maps}(c).

To compare the relative contribution from $\sige$ and $\sigb$ into the resistance measurement, we introduce the edge conductance as $\Gedge$= $\sige/\ell_p$. The fitting parameters $\Gedge$ and $\sigb$ as a function of $V_g$
for HB4 device are shown in Fig.~\ref{fig:maps}(d).
In the gap region corresponding to $|V_g-\Vmax|<1$~V, at $T=3$~K, $\sigb$ is about 5 times smaller than $\Gedge$ and the edge conductance dominates.
%
At the contrary,
$\Gedge$ vanishes outside the band-gap region. This disappearance is also observed in HB1 and HB6\cite{SM}, and is convincing evidence of the topological nature of the edge states.
At the top of the VB, $\Gedge$ is non-zero and even has a local maximum as seen in Fig.~\ref{fig:maps}(d) when $V_g-\Vmax \simeq -1$~V. Actually, we cannot attribute this phenomenon to possible inhomogeneities of the HB4 device, since the similar behavior is also reproduced in HB6 and HB1. Moreover, the non-zero $\Gedge$ contribution in the top VB region is not surprising in view of recent theoretical studies predicting the coexistence of edge and bulk states in complex VB of HgTe~QWs~\cite{q13}.
Note that the valence band of our 3L InAs/GaInSb QWs~\cite{SM} is similar to the ones of HgTe QWs~\cite{2016Krishtopenko}.
Additionally, the inset of Fig.~\ref{fig:maps}(d) shows $\Gedge$ and $\sigb$ at $V_g= \Vmax$ and
confirms the main points of the previous analysis: $\sigb$ has a strong $T$-dependance due to thermal activation above 40~K, while $\Gedge$ dominates the bulk contribution below 40~K.
Clearly, the two curves cross at $T\simeq$~40 K.
Above 40~K, the sudden collapse of $\Gedge$ may indicate a brutal disappearance of the edge states, but a quantitative analysis of $\Gedge$ is difficult in this temperature range.
%
%
\comment{The non-zero edge $\Gedge$ in the VB region is also not surprising. Recent theoretical studies predict the coexistence of edge and bulk states in complex VB of HgTe~QWs~\cite{q13}. We note that the valence band of our 3L InAs/GaInSb QWs~\cite{SM} is similar to the ones of HgTe QWs~\cite{2016Krishtopenko}. The edge state conductivity is resilient to magnetic fields up to 12~T in accordance with the LL calculations (see Figs~\ref{fig:intro}(e,h)), which show the crossing of zero-mode LLs in S3198 (S3052) only at $B\simeq 15$~T (20~T).
}

Assuming that the edge states are helical, the edge conductance in the diffusive regime is given by $\Gedge=(e^2/h)\lambda/\ell_p$, where $\lambda$  is the characteristic length at which the two counter propagating  edge states equilibrate. From $\sige$ at $V_g= \Vmax$ and the lowest temperature, one extracts $\lambda=24$, $2$, $4$ and 10~$\mu$m for HB0, HB1, HB4 and HB6 devices, respectively.
For HB0 device, as $\lambda(HB0)=24$ $\mu$m is larger than the distance $\ell_p$ between the HB0 probes, $\Gedge(HB0)$ goes beyond the quantum limit $e^2/h$, and thus cannot be attributed to topological edge states only: additional parasitic edge conduction is at play. By contrast, in HB1, HB4 and HB6 devices, $\Gedge$ remains smaller than $e^2/h$, and edge conduction via topological edge states only is possible with $\lambda$ values comparable to those of the literature~\cite{2015Du,2016Couedo}. Figure~\ref{fig:maps}(e) summarizes the values of ($\sigb$,~$\Gedge$) at $V_g= \Vmax$ for all the devices.
In HB0 device, $\sigb \simeq \Gedge$, and the current flows equally in bulk and edges. By contrast, in the other devices, the current in the band-gap region flows mainly on the edges at $T= 3$~K.

\emph{Summary.}-- We have demonstrated large inverted band-gap in strained 3L InAs/GaInSb QWs, whose value is comparable with those in compressively strained HgTe QWs~\cite{q7} and 1$T'$-WTe$_2$ monolayers~\cite{q1}. The band-gap in 3L InAs/GaInSb QWs can be even higher than our reported values~\cite{2018KrishtopenkoScience}. Quantitative analysis of the experimental data evidenced topological edge channels.


\begin{acknowledgments}
This work was supported by the Terahertz Occitanie Platform, by CNRS through IRP ``TeraMIR'', by the French Agence Nationale pour la Recherche (Colector project, EquipEx EXTRA), by the European Union through the Marie-Curie grant (agreement No~765426) from Horizon 2020 research and innovation programme.
\end{acknowledgments}


\end{document}



\title{Supplementary materials for `Large inverted band-gap in strained three-layer InAs/GaInSb quantum wells'}


\author{C.~Avogadri}
\affiliation{Laboratoire Charles Coulomb (L2C), UMR 5221 CNRS-Universit\'e de Montpellier, Montpellier, France.}

\author{S.~Gebert}
\affiliation{Laboratoire Charles Coulomb (L2C), UMR 5221 CNRS-Universit\'e de Montpellier, Montpellier, France.}
\affiliation{Institut d'Electronique et des Syst\`emes (IES), UMR 5214 CNRS-Universit\'e de Montpellier, Montpellier, France.}

\author{S.~S.~Krishtopenko}
\affiliation{Laboratoire Charles Coulomb (L2C), UMR 5221 CNRS-Universit\'e de Montpellier, Montpellier, France.}

\author{I.~Castillo}
\affiliation{Laboratoire Charles Coulomb (L2C), UMR 5221 CNRS-Universit\'e de Montpellier, Montpellier, France.}

\author{C.~Consejo}
\affiliation{Laboratoire Charles Coulomb (L2C), UMR 5221 CNRS-Universit\'e de Montpellier, Montpellier, France.}

\author{S.~Ruffenach}
\affiliation{Laboratoire Charles Coulomb (L2C), UMR 5221 CNRS-Universit\'e de Montpellier, Montpellier, France.}

\author{C.~Roblin}
\affiliation{Laboratoire Charles Coulomb (L2C), UMR 5221 CNRS-Universit\'e de Montpellier, Montpellier, France.}

\author{C.~Bray}
\affiliation{Laboratoire Charles Coulomb (L2C), UMR 5221 CNRS-Universit\'e de Montpellier, Montpellier, France.}

\author{Y.~Krupko}
\affiliation{Laboratoire Charles Coulomb (L2C), UMR 5221 CNRS-Universit\'e de Montpellier, Montpellier, France.}

\author{S.~Juillaguet}
\affiliation{Laboratoire Charles Coulomb (L2C), UMR 5221 CNRS-Universit\'e de Montpellier, Montpellier, France.}

\author{S.~Contreras}
\affiliation{Laboratoire Charles Coulomb (L2C), UMR 5221 CNRS-Universit\'e de Montpellier, Montpellier, France.}

\author{A.~Wolf}
\affiliation{Technische Physik, Physikalisches Institut and W\"urzburg‐Dresden Cluster of Excellence ct.qmat, Am Hubland, D-97074 W\"urzburg, Germany}

\author{F.~Hartmann}
\affiliation{Technische Physik, Physikalisches Institut and W\"urzburg‐Dresden Cluster of Excellence ct.qmat, Am Hubland, D-97074 W\"urzburg, Germany}

\author{S.~H\"ofling}
\affiliation{Technische Physik, Physikalisches Institut and W\"urzburg‐Dresden Cluster of Excellence ct.qmat, Am Hubland, D-97074 W\"urzburg, Germany}

\author{G. Boissier}
\affiliation{Institut d'Electronique et des Syst\`emes (IES), UMR 5214 CNRS-Universit\'e de Montpellier, Montpellier, France.}

\author{J.-B. Rodriguez}
\affiliation{Institut d'Electronique et des Syst\`emes (IES), UMR 5214 CNRS-Universit\'e de Montpellier, Montpellier, France.}

\author{S.~Nanot}
\affiliation{Laboratoire Charles Coulomb (L2C), UMR 5221 CNRS-Universit\'e de Montpellier, Montpellier, France.}

\author{E.~Tourni\'e}
\affiliation{Institut d'Electronique et des Syst\`emes (IES), UMR 5214 CNRS-Universit\'e de Montpellier, Montpellier, France.}

\author{F.~Teppe}
\email[]{frederic.teppe@umontpellier.fr}
\affiliation{Laboratoire Charles Coulomb (L2C), UMR 5221 CNRS-Universit\'e de Montpellier, Montpellier, France.}

\author{B.~Jouault}
\email[]{benoit.jouault@umontpellier.fr}
\affiliation{Laboratoire Charles Coulomb (L2C), UMR 5221 CNRS-Universit\'e de Montpellier, Montpellier, France.}

\maketitle

\renewcommand{\thefigure}{S\arabic{figure}} %
\renewcommand{\thetable}{S\arabic{table}}   %
\renewcommand{\theequation}{S\arabic{equation}}   %

\subsection{Band structure and growth scheme of the samples}
Band structure calculations were performed by using the eight-band Kane model~\cite{refSM1}, which directly takes into account the interactions between $\Gamma_6$, $\Gamma_8$, and $\Gamma_7$ bands in bulk materials. This model well describes the electronic states in a wide range of narrow-gap semiconductor QWs, particularly in the broken-gap InAs/Ga(In)Sb quantum wells (QWs)~\cite{refSM2,refSM3,refSM4,refSM5}. In the eight-band Kane Hamiltonian, we also took into account the terms, describing the strain effect arising because of the mismatch of lattice constants in the buffer, QW layers, and AlSb barriers. To calculate LLs, we used the so-called axial approximation. Within this approximation, one keeps in-plane rotation symmetry by omitting the warping terms in the Hamiltonian. The calculations had been performed by expanding the eight-component envelope wave functions in the basis set of plane waves and by numerical solution of the eigenvalue problem. Details of calculations, notations of the LLs, and the form of the Hamiltonian can be found elsewhere~\cite{refSM1}. Parameters for the bulk materials and valence band offsets used in the eight-band Kane model are taken from Refs~\cite{refSM6,refSM7,refSM8}.

\begin{figure*}[!ht]
\includegraphics[width=1.0\linewidth]{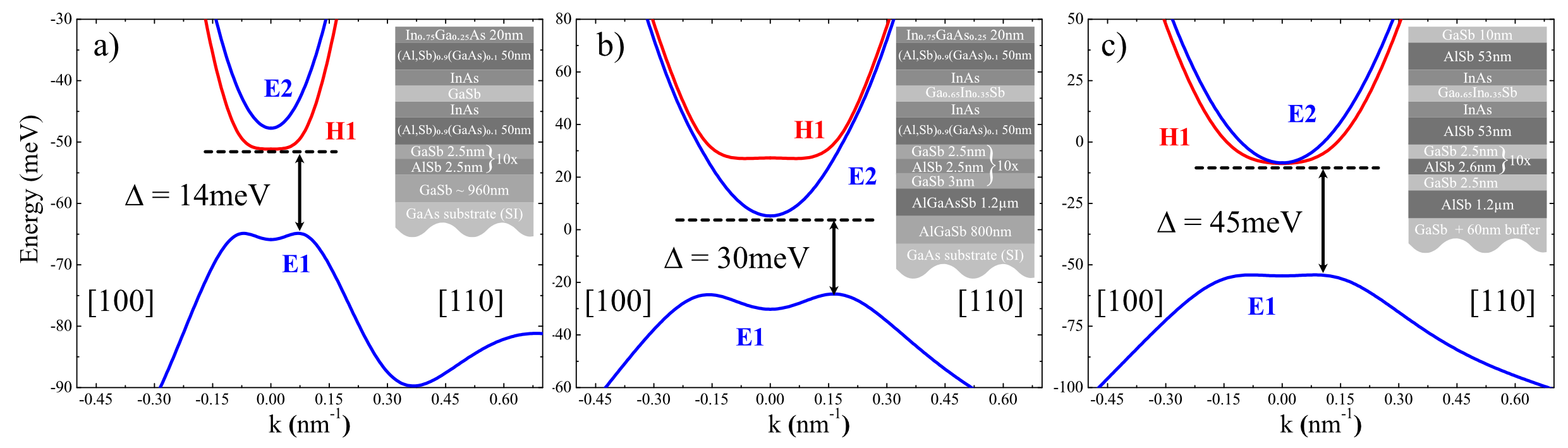}
\caption{\label{Fig:Bands} Band structure of three samples: S3054 (a), S3052 (b) and S3198 (c). Numbers in brackets denote crystal directions. The values of $\Delta$ represent the theoretical band-gap. The insets show the sequence and width of the grown layers for each sample.}
\end{figure*}

Figure~\ref{Fig:Bands} provides the band structure and the growth scheme for all three samples under study (see the main text). Blue and red curves represent the band dispersion of electron-like and hole-like subbands, respectively. The electronic levels have been classified as electron-like or hole-like subbands by comparing the relative contribution to this level at zero quasimomentum from the basis states of $|\Gamma_6,\pm1/2\rangle$, $|\Gamma_7,\pm1/2\rangle$ and $|\Gamma_8,\pm1/2\rangle$ bands with the contribution from the heavy-hole band $|\Gamma_8,\pm3/2\rangle$~\cite{refSM9}.

\subsection{Activation energy}
To determine experimental values of the band-gap, we have focused on measuring the longitudinal resistivity $\rho_{xx} \simeq R_{14,23}W/l_p$ in local geometry. Here, $W$ is the width of fabricated Hall bar, while $l_p$ is the distance between the lateral probes (see Fig.~1 in the main text). Note that $\rho_{xx}$ also includes the edge state contribution. Similar to the case of InAs/GaInSb QW bilayers~\cite{refSM10}, the temperature dependance of the resistivity were fitted by a sum of three terms:
\begin{equation}
\label{eq:1}
\left(\rho_{xx}^\mathrm{peak}\right)^{-1}=
\sigma_a
\exp{
\left(
-
\frac{\Delta}{2k_BT}
\right)
}
+ \sigma_\mathrm{l}(T)
+\sigma_0.
\end{equation}
Here, in the right hand side, the first term corresponds to the activation energy, the second term is due to localization gap, and the last term $G_0$ represents the conductance of the edge states.

We assume that the localization is induced by nearest neighbor hopping and takes the form:
\begin{equation}
\sigma_\mathrm{l}(T)=
\sigma_\mathrm{loc}
\exp{
\left(
-\frac{\Delta_\mathrm{loc}}{k_B T}
\right)}.
\end{equation}
Two Hall bars, namely HB4 and HB6 devices, were fabricated from sample S3198. Starting from $T=300$ mK, the devices were progressively warmed up, while the gate voltage was continuously swept. Then, the maximum of the resistivity was retrieved for each temperature.

\begin{figure*}[h!]
\begin{minipage}{0.495\linewidth}
\includegraphics[width=1.0\linewidth]{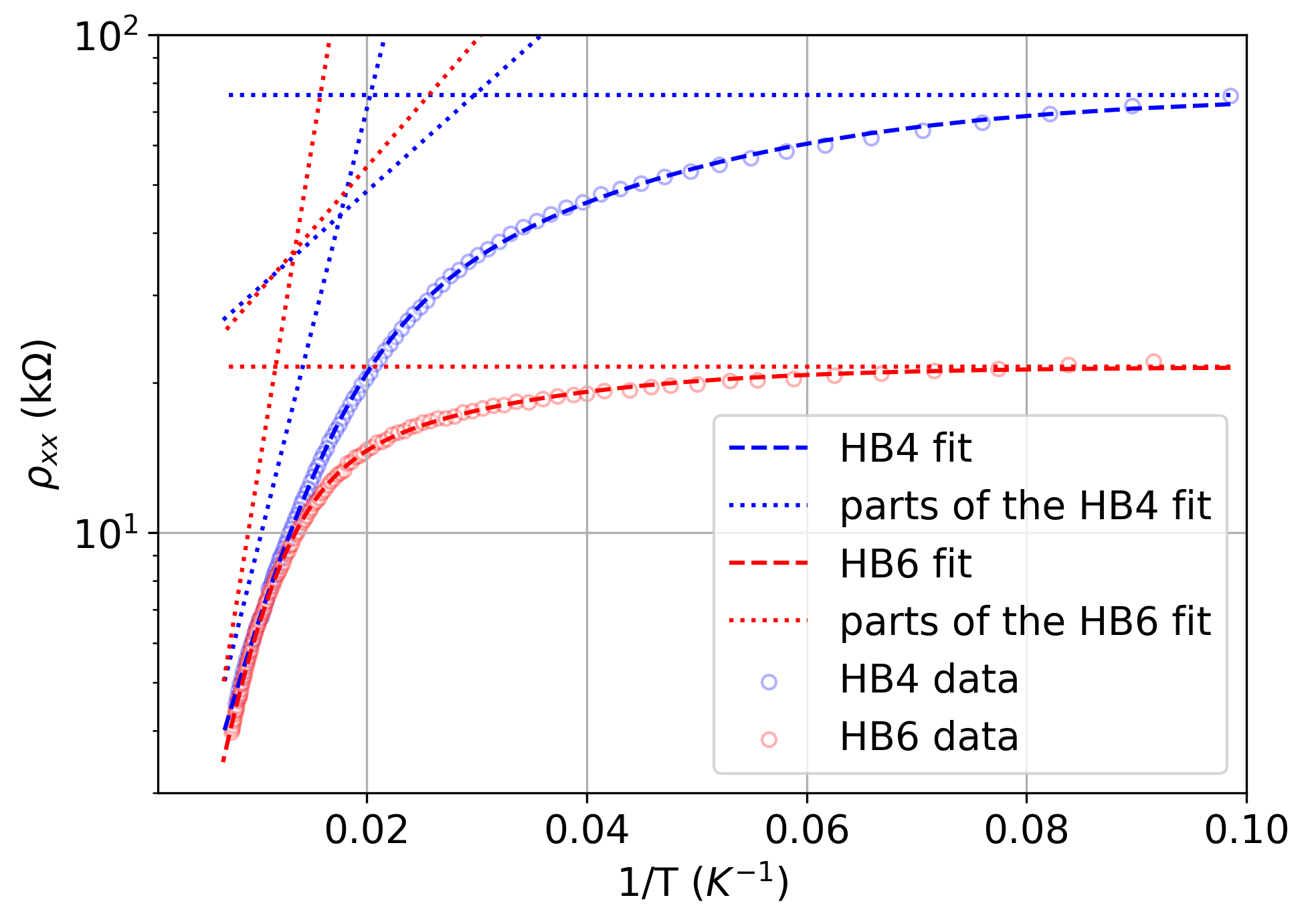}
\end{minipage}
\begin{minipage}{0.495\linewidth}
\includegraphics[width=1.0\linewidth]{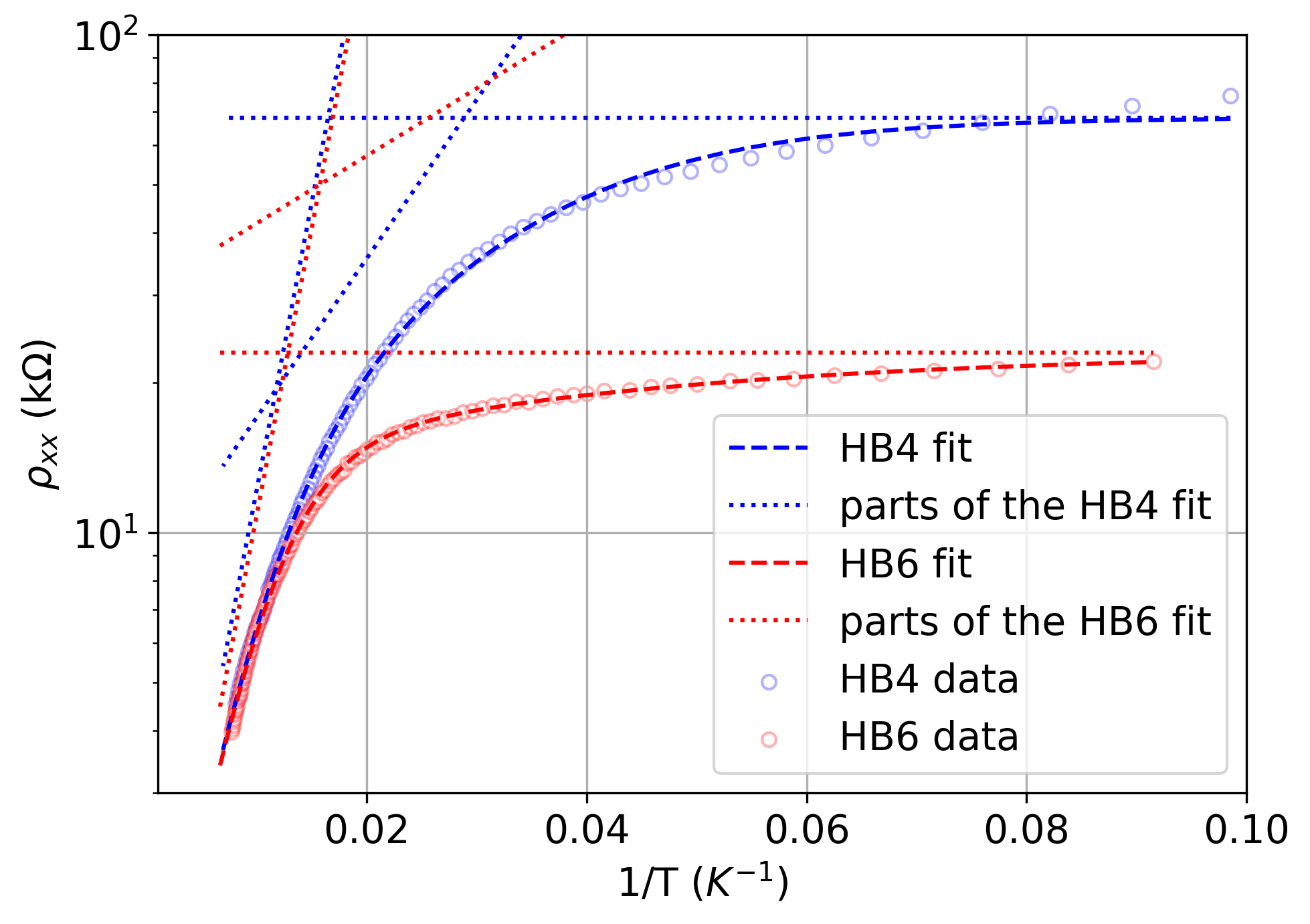}
\end{minipage}
\caption{\label{fig:fit1} Fits of $\rho_{xx}^{\mathrm{peak}}$ as a function of temperature for HB4 and HB6 devices fabricated from sample S3198 (see the main text). The left panel corresponds to the fits performed separately for HB4 and HB6 devices. The right panel represents the fits, which were are done simultaneously imposing the same activation energy $\Delta$ for both devices.}
\end{figure*}

\begin{figure}
	\includegraphics[width=0.5\linewidth]{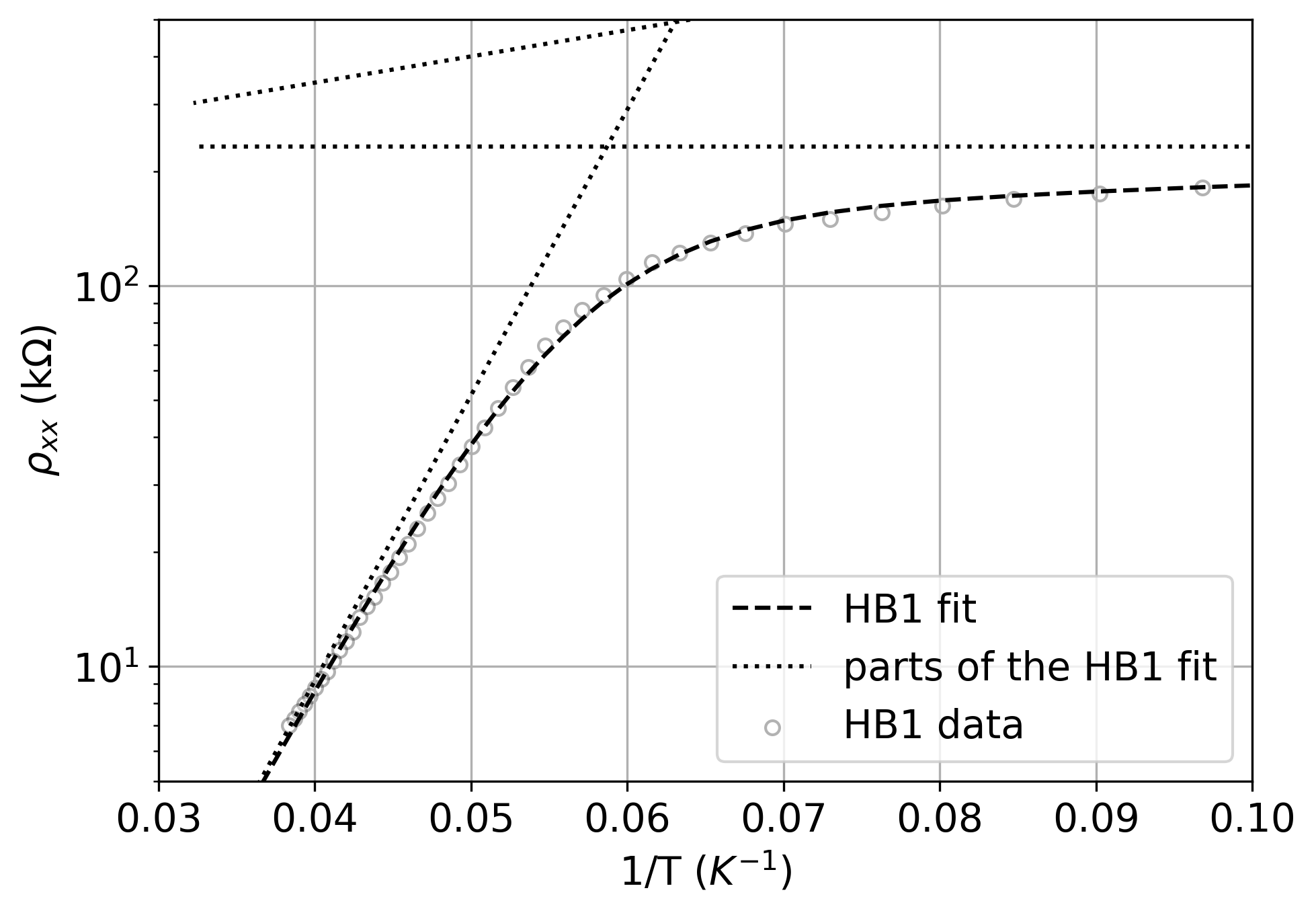}
	\caption{Fits of $\rho_{xx}^{\mathrm{peak}}$ as a function of temperature for HB1 device fabricated from sample S3052 (see the main text).}
	\label{fig:fit3}
\end{figure}

To extract the band-gap for sample S3198, we have first fitted separately the temperature dependence of the resistivity peak measured in HB4 and HB6 devices (see Fig.~\ref{fig:fit1}).
The fits give $\Delta=$ 36~meV and $\Delta= $54~meV for HB4 and HB6 devices, respectively.
Then, we also fitted simultaneously $\rho_{xx}^{\mathrm{peak}}$ for both devices, imposing the same values of $\Delta$. The latter also gives reasonable fits with $\Delta$ = 46~meV. By combining experimental band-gap values obtained by different types of the fitting for both devices, we have got $\Delta= 45 \pm 8$~meV, which also includes the device to device variations. For sample S3052, we have performed the fit of experimental data obtained in the HB1 device only (see the main text). The device was progressively warmed up, while the resistance was measured at the gate voltage corresponding to the resistivity peak at $T= 300$~mK. The fitting analysis provided in Fig.~\ref{fig:fit3} gives $\Delta= 30\pm 0.3$~meV. Note that the error bar obtained in HB1 device includes only the standard deviation from the data fitting.

By analyzing the transport measurements in HB6 device, one can also estimate experimental values of the density of states $D$ inside the band-gap. The $D$ values is defined by the difference $\Delta{V}$ between the gate voltage values corresponding to the middle of the band-gap and the van Hove singularity observed in HB6 device (see the main text):
\begin{equation}
\Delta = 2 \Delta{V}(C_g/q) /D,
\end{equation}
where $-q$ is the electron charge, $C_g$ is the gate capacitance. With $\Delta V \simeq 2.5$~V, $\Delta= 45$~meV, we get $D= 8\times 10^{12}$~eV$^{-1}$cm$^{-2}$. The latter is approximately ten times smaller than the density of states expected in the conduction and valence bands (see Fig.~1(d) in the main text), thus confirming the existence of the band-gap.

\begin{figure}[h!]
\includegraphics[width=0.5\linewidth]{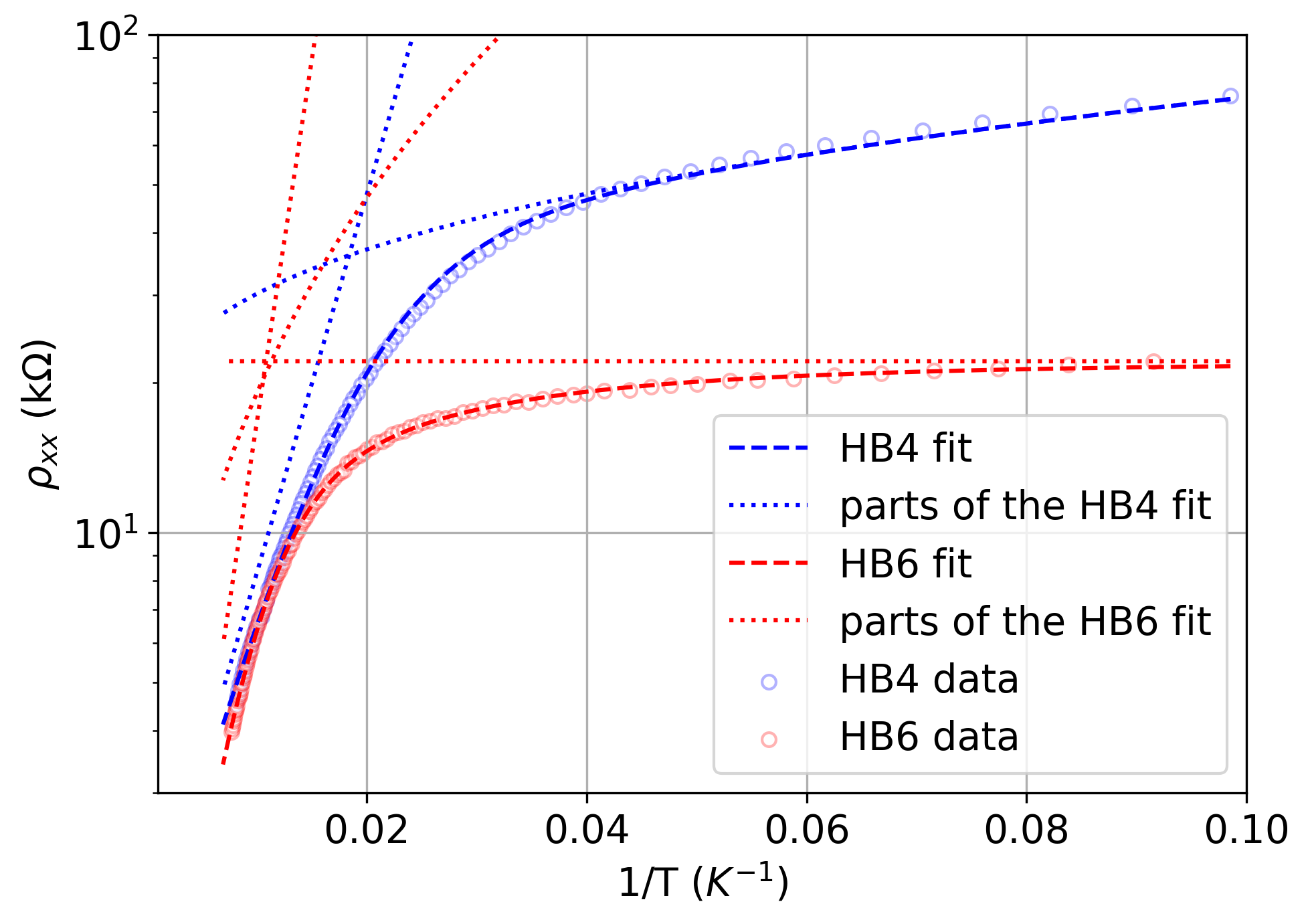}
\caption{Fits of $\rho_{xx}^{\mathrm{peak}}$ for HB4 and HB6, assuming variable range hopping for the localization contribution $\sigma_\mathrm{l}(T)$ (see Eq.~(\ref{eq:VHS})). The fits were performed separately for HB4 and HB6 devices.}
\label{fig:fit4}
\end{figure}

Additionally, let us detail Mott's approach and nearest neighbor hopping. In this model, the energy barrier between nearest sites separated by a distance $a$ is defined by the energy gap:
\begin{equation}
\Delta_\mathrm{loc}= (D a^2)^{-1}.
\end{equation}
For three devices HB1, HB4, and HB6, one cand find $\Delta_\mathrm{loc}= 3.4 \pm 1.6$~meV resulting in $a \sim $50--100~nm, a rather large value.
We note that, strictly speaking, the localization contribution $\sigma_\mathrm{l}(T)$ in Eq.~(\ref{eq:1}) can be caused by a mechanism other than the nearest neighbor hopping. For instance, Mott's variable range hopping (VRH) follows the expression:
\begin{equation}
\label{eq:VHS}
\sigma_\mathrm{l}(T)=\sigma_\mathrm{loc}\exp{\left[-\left(\frac{\Delta_\mathrm{loc}}{k_B T}\right)^{1/3}\right]}.
\end{equation}
Being introduced in Eq.~(\ref{eq:1}), it also gives very good fits as seen from Fig.~\ref{fig:fit4}. The gap estimation remains identical for S3052 -- $\Delta_{VRH}(S3052)=30$~meV, and slightly differ for V3198 -- $\Delta (S3198)= 41 \pm 12$~meV.
For the sake of clarity, we stick to the simplest NN localization model in the main text.

\subsection{Square lattice model of conductances}
In the main text, each of the Hall bar devices is modeled by 2D $n_L \times n_W$ square-lattice network of resistances (see Fig.~\ref{fig:model}). On the two edges of the Hall bar,
the conductance is given by $\sigma_\mathrm{edge}$ (in units of length per Ohm).
Thus, the conductance of a single edge resistance is given by $\sigma_\mathrm{edge}/l_c$,
where $l_c$ is the side length of the squares. In the bulk, the conductance is defined to the bulk conductivity (in ohm per square).  Thus, the conductance of one of the inner resistances is given by $\sigma_\mathrm{bulk}$. At each $i$ node, the potential $V_i$ obeys the equation of current conservation:

\begin{figure}
	\includegraphics[width=0.45\linewidth]{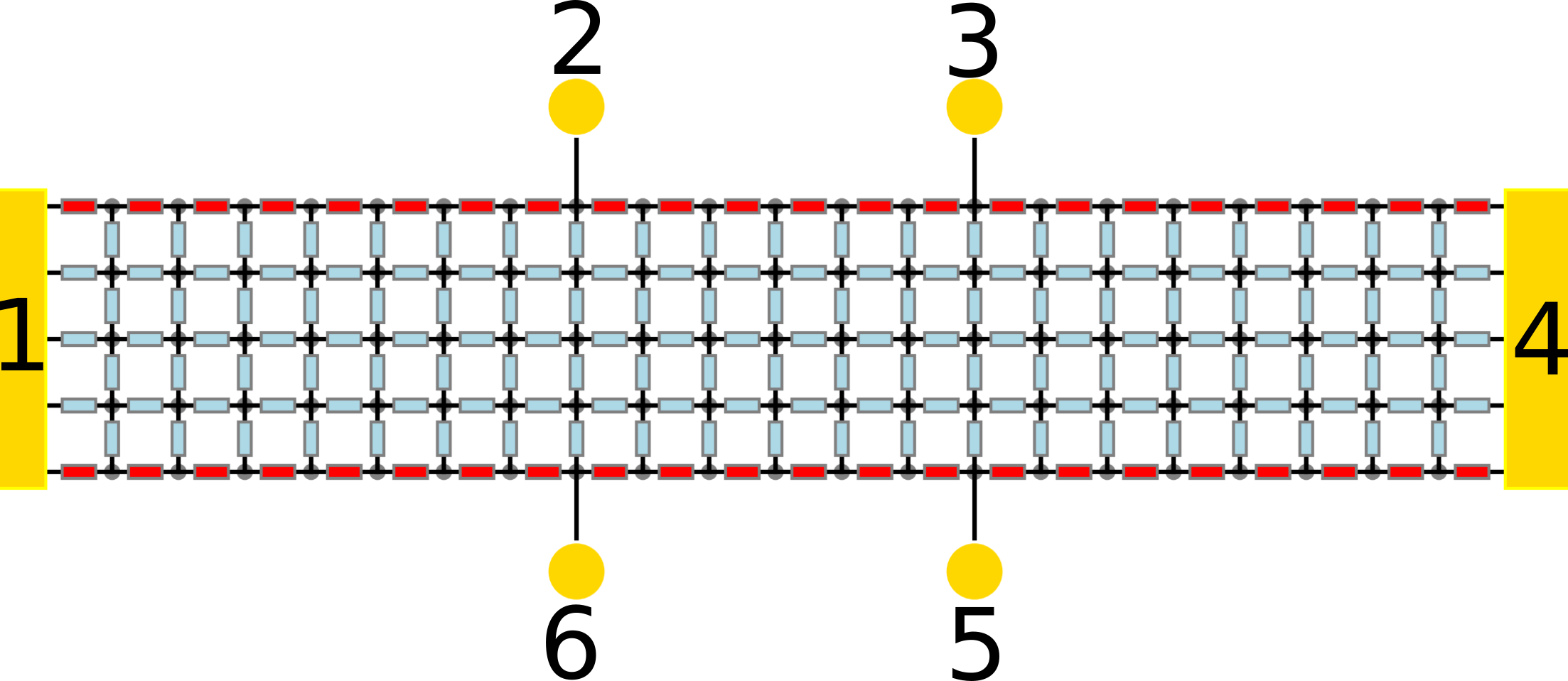}
	\caption{Sketch of resistance square network for a given Hall bar geometry ($n_L=22$, $n_W=4$). The squares have a side length $l_c$. The red resistors represent the contribution from 1D edge state. They have a resistance $\sigma_\mathrm{edge}^{-1} l_c$. The light blue resistors model the 2D bulk and they have a resistance $\sigma_\mathrm{bulk}^{-1}/\square$.}
	\label{fig:model}
\end{figure}

\begin{equation}
\label{eq:Cond}
\sum_{j=\mathrm{NN}} G_{i,j} (V_j-V_i)= B_i,
\end{equation}
where $j$ lists the nearest neighbors, $G_{i,j}$ is the conductance element between the sites $i$ and $j$ (which equals either $\sigma_\mathrm{edge}/l_c$ or $\sigma_\mathrm{bulk})$. Note that $B_i$ in Eq.~(\ref{eq:Cond}) is defined by the boundary condition. The $B_i$ value is non-zero only if the $i$ node is connected to an additional contact acting as a current source or drain. In the latter case, $B_i= G_c (V_i-V_c)$, where $G_c$ is the contact resistance, $V_c$ is the voltage of the contact. The exact value of $G_c$ is unimportant for the measurements in the four-probe geometry. Obviously, the network is described by the set of coupled linear equations for the unknown $V_i$'s. This problem can be solved numerically when the voltages of the source and drain contacts  are imposed.

\begin{figure}[ht!]
	\includegraphics[width=0.6\linewidth]{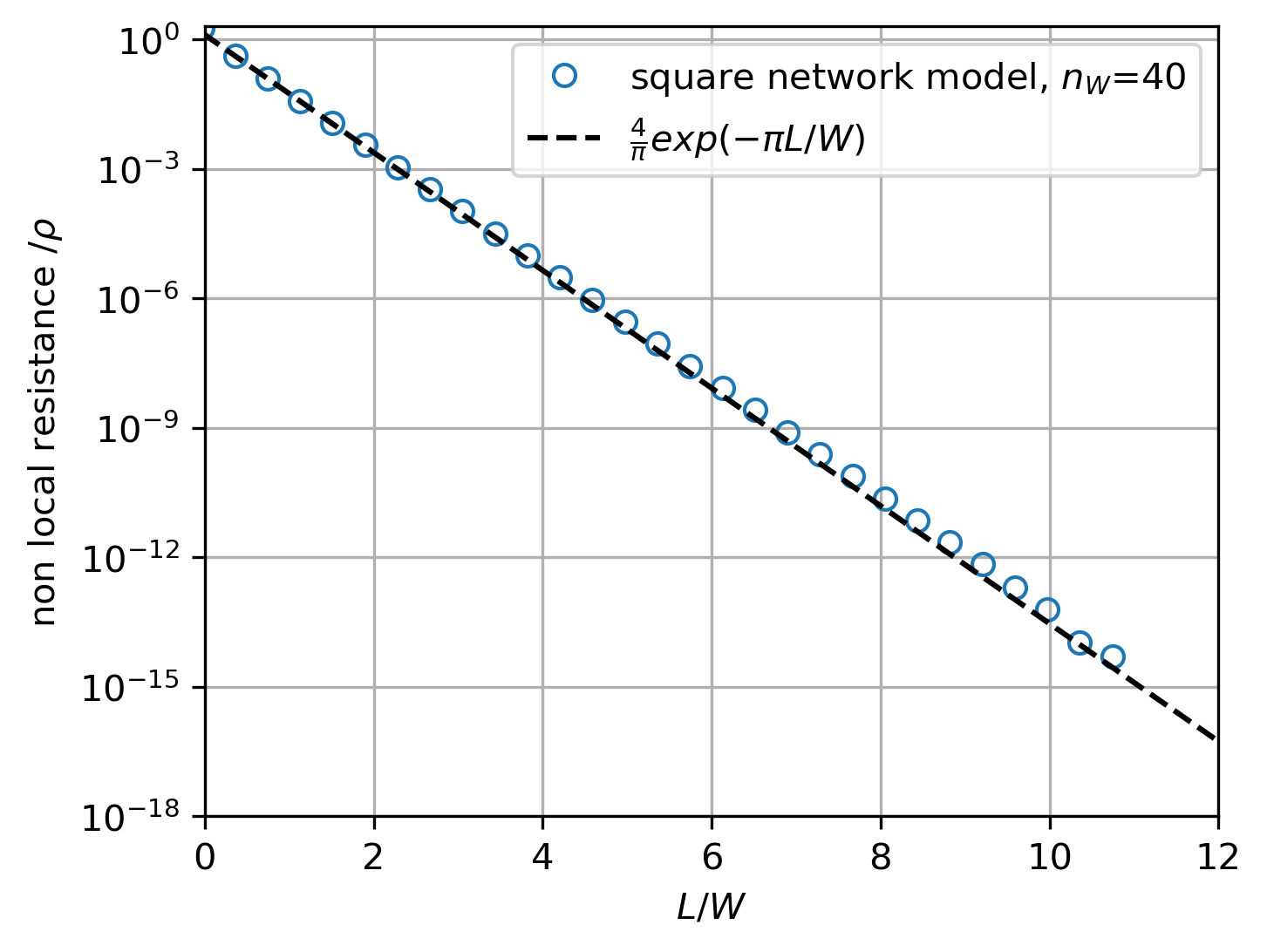}
	\caption{Comparison of the numerical estimate of the non-local resistance, $R_{25,34}$, as given by the square network, and conventional formula $R_{25,34}=4/\pi\rho \exp(-{\pi}L/W)$. Here, $L$ is the distance between the probes $2$ and $3$, while $W$ is the Hall bar width. The width of the Hall bar has been divided into $n_W=40$ squares.}
	\label{fig:flow}
\end{figure}

When $n_L, n_W \rightarrow \infty$, the square network model becomes equivalent to the finite difference method
for the resolution of the Laplace equation $\nabla^{2} V=0$.
The voltages of the source and drain contacts are imposed, yielding Dirichlet boundary conditions.
The boundary conditions imposed
by the edge states are:
\begin{equation}
\sigma_\mathrm{edge} \partial^2 V/ \partial x^2 +
\sigma_\mathrm{bulk} \partial V / \partial y=0.
\end{equation}
We have checked that the model reproduces satisfactorily the expected non-local resistance $R_{25,34}$ of an homogeneous device (i.e. without edge conduction), when the size of the square mesh $l_c$ tends to zero. In practice, it is sufficient to choose $n_W \ge 40$. This is illustrated in Fig.~\ref{fig:flow}, where the non-local resistance of the square network has been calculated when the ratio $L/W$ of the Hall bar is varied. Here, $L$ is the distance between the contacts $2$ and $3$, $W$ is the Hall bar width.  The model reproduces satisfactorily the well known formula for the current spreading:
\begin{equation}
R_{25,34}= \frac{4}{\pi}\rho \exp(-\pi L/W),
\end{equation}
up to the relative precision of $10^{-14}$. All the calculation presented in the manuscript have been done with $n_W \ge 20$.

\begin{figure}[ht!]
	\includegraphics[width=0.75\linewidth]{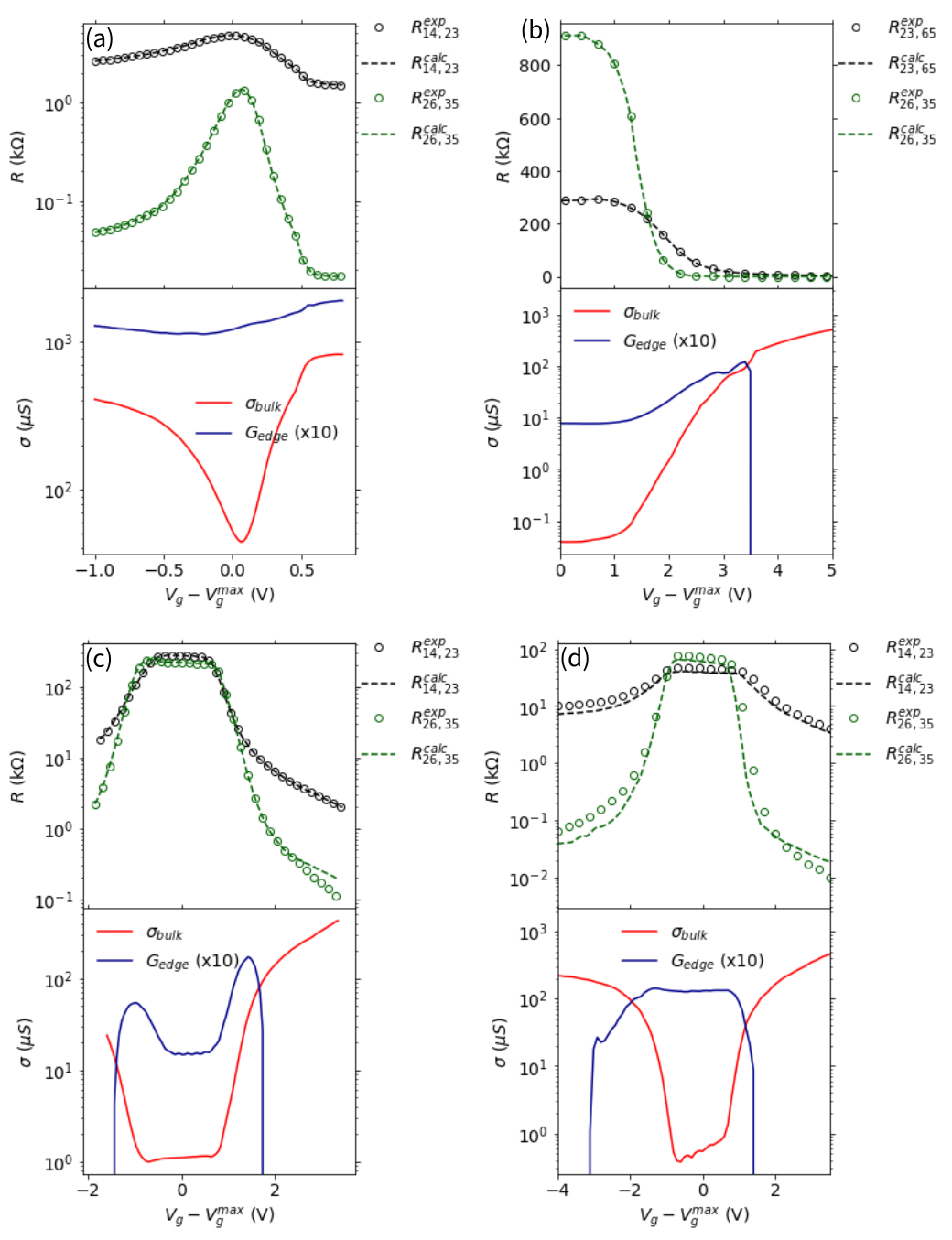}
	\caption{(a) upper figure: experimental non-local and local resistances for HB0 device (open circles) and their values obtained by fitting analysis in the resistive network model (dashed lines). Lower figure: fitting parameters $G_\mathrm{edge}= \sigma_\mathrm{edge}/\ell_p$
	and $\sigma_\mathrm{bulk}$. (b--d): same figures for HB1, HB4 and HB6 devices, respectively. }
	\label{fig:sqlatfit}
\end{figure}

In the main text, resistance square networks have been applied to each Hall bar device, and both bulk and edge contributions were extracted by fitting the non-local and local transport measurements. The agreement between the experimental and calculated resistances is demonstrated in Fig.~\ref{fig:sqlatfit}. For HB0 device, whose data are provided in panel (a), the edge contribution does not vanishes when the Fermi level lies in the bulk bands, suggesting parasitic edge conduction. By contrast, for HB1 (panel (b)), HB4 (panel (c)) and HB6 (panel (d)) devices,
$G_\mathrm{edge}\ge\sigma_\mathrm{bulk}$ in the band-gap but $G_\mathrm{edge}\le\sigma_\mathrm{bulk}$ in the bulk bands. This behavior suggests that helical edge conduction is at play.
As it is seen, an unexpected overshoot of the edge conductance appears at the boundary between the band-gap and the bulk bands. Experimentally, this is caused by the fact that the width of the non-local resistance peak was systematically larger than the width of the local resistance peak in our measurements.
In the gap, the observed local resistance of the HB4 device is considerably larger in Fig.~\ref{fig:sqlatfit} than in Fig.~3 of the main text. The two figures correspond to two different cool-downs of the same device. This illustrates the importance of mastering the electrostatics of these devices in future works.

\begin{figure}
	\includegraphics[width=0.65\linewidth]{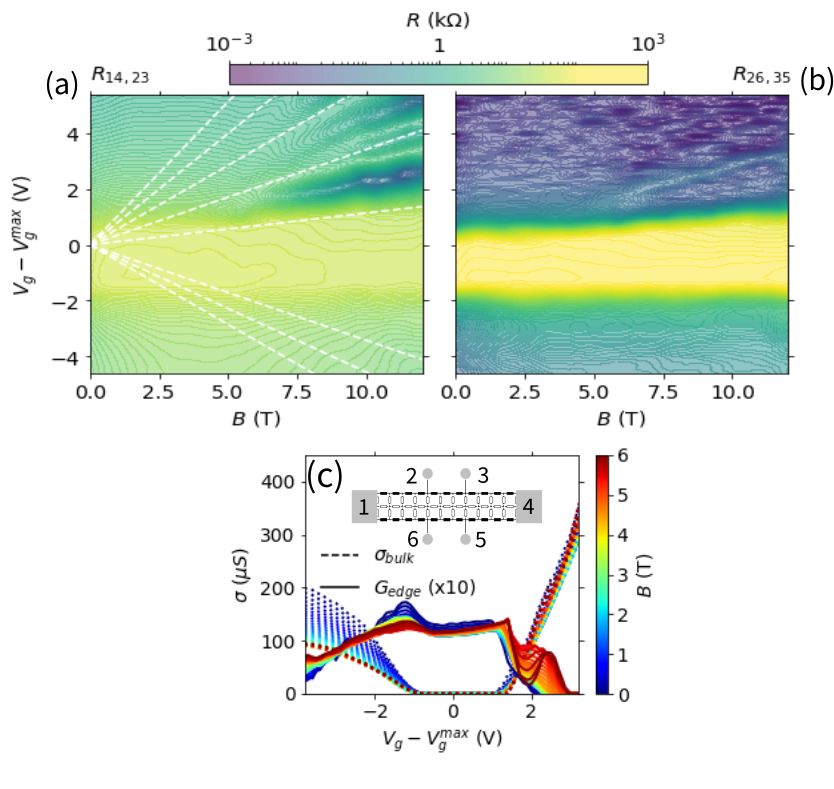}
	\caption{(a) Colormap of the local resistance $R_{14,23}$, as a function of $B$ and $V_g$ at $T=300$~mK. The white lines are guides for the eyes, corresponding to integer filling factors $\nu=-5,-4,-3,1,3,5,7$ and 9. (b)~Similar colormap for the non local resistance $R_{26,35}$. The arrow shows the non local resistance at $\nu \simeq 2.5$ induced by backscattering of the topmost LL~\cite{1990McEuen}. (c)~Fitting parameters $\sigb$ and $\Gedge$, as a function of $V_g$, at different magnetic fields. The inset sketches a minimalist model of a square resistor network, where $\sigb^{-1}$ and $\sige^{-1}l_c$ are indicated as white and black resistors respectively ($l_c$ is the size of the square).}
	\label{fig:maps}
\end{figure}

\subsection{Magnetoresistance}
Fig. S8(a) shows  colormaps of the local magnetoresistance $R_{14,23}$ for HB6 at $T= 300$~mK,
as a function of $B$ and $V_g$.
Landau levels (LL) of the CB are well visible on $R_{14,23}$ for $B > 5$~T, $V_g-\Vmax>1$ V.
Some LLs are also visible in the VB.
The LL formation is evidenced by the white lines which correspond to filling factor $\nu=$
-5,-4,-3,  1, 3, 5, 7, and 9.
At $V_g - \Vmax = 5$~V,
the SdH concentration $n_{SdH}= 1.06 \times 10^{11}$ cm$^{-2}$
is in accordance with the Hall concentration $n_H= 1.15 \times 10^{11}$~cm$^{-2}$.
At $B = 12$~T and $\nu=2$, the device is in the quantum Hall regime and
$R_{14,23}$ becomes as low as $\simeq 20~\Omega$,
which gives a residual conductivity $\sigma_{xx} \simeq 0.1~\mu$S.
Thus, parallel conduction is negligible in HB6.
%

Fig.~S8(b) shows a similar colormap for the nonlocal resistance $R_{26,35}$.
%
%
%
In the conduction and valence band,
the non-local resistance finds its origin
in the current spreading in the bulk of the Hall bars:
$R_{26,35}^\mathrm{exp} \simeq  (4\rho/\pi) \exp (-\pi \ell_p/W)$
where $\rho= R_{14,23}^\mathrm{exp} W/\ell_p$ is the bulk resistivity,
$\ell_p$ is the distance between probes 2 and 3 and $W$ is the Hall bar width.
This yields, $R_{26,35}^\mathrm{exp} \simeq R_{14,23}^\mathrm{exp}\times 10^{-2}$,
as approximately observed when $|V_g - V_g^\mathrm{max}|>2$, see Figs.~S8(a,b).
Remarkably, the non local resistance has a maximum along $\nu \simeq 2.5$,
whose relatively modest amplitude (200 $\Omega$) is well explained
by the backscattering of the topmost LL (BTLL)~\cite{1990McEuen}.
The fitting parameters $\Gedge$ and
$\sigb$  are shown in Fig.~S8c for HB6,
as a function of $V_g$,
for non-quantizing magnetic fields.
%
%
In the gap ($|V_g - \Vmax| < 1 $ V),
$\sigb \ll \Gedge$
and edge conductance dominates. Numerically,
$1/\sigb$ reaches values as high as 1 M$\Omega$,
whereas the edge conductance is about 10 $\mu$S, at least ten times higher than $\sigb$.
In the VB, $\sigb$ decreases with $B$, as expected in the two-fluid model,
whereas $\Gedge$ decreases smoothly when $V_g$ decreases.
In the CB,
the decrease of $\Gedge(V_g)$ is much steeper and
$\sigb$ gives the main contribution to both local and non local resistances.
Even if our model is only adapted to helical edge states,
it correctly interprets
the BTLL~\cite{1990McEuen} non-local resistance as an additional edge contribution,
visible as a peak around $V_g-\Vmax \simeq 2$~V.

We also note that the edge state conductivity is resilient to magnetic fields up to 12~T in accordance with the LL calculations shown in Fig.~1(e,h) in the main text, which show the crossing of zero-mode LLs in S3198 (S3052) only at $B\simeq 15$~T (20~T).
